\documentclass[10pt,dvips]{article}
\usepackage{epsfig,times}

\begin{document}

\begin{center}
{\bf \Large A search of multiparticle correlations in  10.7 A GeV $^{197}Au$ and 
200 A GeV $^{32}S$  interactions with emulsion nuclei by the Hurst method} \\ [2mm]
{\it A.Sh.Gaitinov, I.A.Lebedev, P.B.Kharchevnikov, V.I.Skorobogatova, A.T.Temiraliev} \\ [2mm]
{Institute of Physics and Technology, Almaty, Kazakhstan}
\end{center}

\begin{abstract}
An analysis of pseudorapidity correlations in 10.7 A Gev $^{197}Au$ and
200 A GeV $^{32}S$ interactions with emulsion nuclei by the normalized range
method has been carried out.
The evidence for events with large multiparticle correlations is presented.
The most significant "correlation force" effect is apparent in the interactions
of light emulsion nuclei (CNO  group) and gold nuclei at an energy of 10.7 A
GeV, corresponding to an absolute disintegration of the target nucleus
$n_b+n_g\simeq 0$.
\end{abstract}

\section {Introduction}

A study of interactions of relativistic nuclei is related to a search
for unusual nuclear matter conditions (as quark-gluon plasma (QGP), for
example), which can be observed at high energies  \cite{drem}.
The behaviour of simple features of these processes, such as total cross
sections, multiplicity and pseudorapidity distributions, etc. is rather
precisely described by various models, even if they are based on the
various
assumptions about  nuclear structure and dynamics of nuclear
interactions. More
detailed information, critical to a choice between existing models, can
be
deduced by investigating a fluctuation structure of secondary particle
distributions.

Recent studies of pseudorapidity distributions have shown that some
considerable deviations from average distribution can be observed in
individual
events. Certainly, these deviations can be initiated by the
statistical factors, connected with finite number of particles in the
given
event. If the statistical explanation cannot describe these
fluctuations, there
are the some dynamic mechanisms, which could produce peculiarities of
pseudorapidity distributions.

In our previous work \cite{leb} we have proposed a normalized range
method for
analysing pseudorapidity correlations in multiparticle production
processes.
This method allows not only to discriminate dynamic correlations from
statistical ones, but also to determine a "force" and "length" of these
correlations.

In the present paper we apply this method to the analysis of
experimental
pseudorapidity distributions of secondary particles obtained in 10.7 A
GeV
$ ^{197} Au$ and 200 A GeV $ ^{32}S$ interactions with emulsion nuclei.

\section{Analysis procedure}

As result of high-energy interaction of two nuclei plenty of secondary
particles is produced. According to the existing notions,
secondary particles, which are "emited" from "interaction volume", have
pseudorapidities, corresponding to a central region of pseudorapidity
distribution. At borders of the distribution fragments (of the target-nucleus
and projectile-nucleus) bring in considerable contribution. And so, in order
to investigate of pseudorapidity
correlations in distribution of particles from "interaction volume"
we have chosen pseudorapidity interval $\Delta \eta =  4$
(for the $ ^{32}S$, 200 A GeV, $ \eta_{min}= 1.0$, $\eta_{max}= 5.0$
and for the $ ^{197}Au$, 10.7 A GeV,  $ \eta_{min}= 0.3$,
$\eta_{max}= 4.3$).
This interval has been subdivided into $k$ parts with $\delta\eta=\Delta\eta /k$.
By counting the number of particles in each subinterval we arrive at a
sequence $n_i^e$.
A pseudorapidity fluctuation, or the normalized relative deviation
of an individual event from average pseudorapidity distribution
\footnote{It is possible to use also absolute deviation
$\xi_i =  n_i^e/n^e - n_i/n.$} is given by
\begin{equation}
\xi_i = \frac{n_i^e/n^e - n_i/n}{n_i/n}= \frac{n_i^e}{n^e} \;
\frac{n}{n_i} - 1
\end{equation}
where $ n_i^e$ is the number of particles in the i-th bin of an event
with
particles number $n^e$, and $n_i= \sum_en_i^e$ is the total number of
particles
for all events in the i-th bin, and $n= \sum_en^e$ is the total number
of
particles for all events.

For analysis we have selected events with the number of secondary
particles (except fragments) greater than 90, because high statistics
are necessary for a study of correlations by our method.

As described in our previous work \cite{leb} for an investigation of
pseudorapidity correlations we analysed the normalized range
$H(k')= R(k')/S(k')$ (where $R(k')$ and $S(k')$ are a "range" and a standard
deviation, which is calculated by Eqs.(\ref{s})-(\ref{1}), see below)
versus the size of the pseudorapidity interval $d\eta =  k'\delta\eta$,
($1\le k' < k$) using a function
\begin{equation}
H(k')= (a k' )^h
\label{6}
\end{equation}
where $a$ and $h$ are two free parameters and $h$ is the correlation
index (or Hurst index). If the signal $\xi_i$ represents white noise
(a completely uncorrelated signal), then $h= 0.5$.
If $h>0.5$, long-range correlations are in a system \cite{h,f}.

In our calculations we have used $k= 8192$. The choice of such large value
of $k$ is not necessary (it is possible using lesser $k$ also). But
the more $k$ the more accuracy of method is approachable.

So, for the sequence $\xi_i$, $1\le i\le k$,
quantities of $R(k)$ and $S(k)$ were calculated by following formulas:
\begin{equation}
S(k)= \left[ \frac{1}{k} \sum^{k}_{i= 1} [\xi_i-<\xi >]^2\right]^{1/2}
\label{s}
\end{equation}
\begin{equation}
R(k)= \underbrace{max X(m,k)}_{1\le m\le k} -\underbrace{min
X(m,k)}_{1\le m\le k}
\end{equation}
where the quantity $X(m,k)$ characterizes the accumulated deviation from
the average
\begin{equation}
< \xi > = \frac{1}{k} \sum^{k}_{i= 1}\xi_i
%\label{1}
\end{equation}
for a certain interval $m\delta\eta$,
\begin{equation}
X(m,k)= \sum^{m}_{i= 1} [\xi_i-<\xi >], \quad 1\le i \le m \le k
\label{1}
\end{equation}

Then the sequence $\xi_i $ has been subdivided into two parts: $\xi_i'$,
$1\le
i\le k'= k/2$ and $\xi_i''$, $k'+1\le i\le k$, and the value of
$H(k/2)= R(k/2)/S(k/2)$ was found for each of the two independent
series.
Similarly $\xi_i'$ and $\xi_i''$ have been subdivided further, followed
by the
calculation of $H(k/4)= R(k/4)/S(k/4)$. This subdivision and analysis
procedure
for newly obtained series of $\xi_i$-values is continued until $d\eta >
0.1$.
$H$ corresponding to the same value of $k'$ have been averaged and drawn
on a
log-log scale as a function of $k'$~(Fig.1).

\begin{figure}[tbh]
\begin{center}
\includegraphics*[width=0.48\textwidth,angle=0,clip]{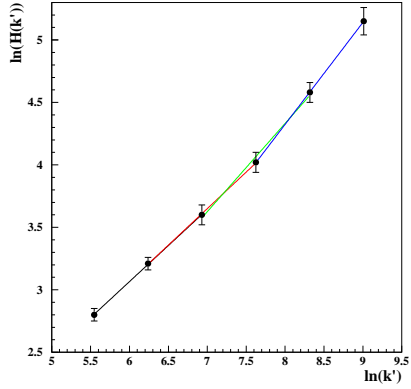}
\caption{\label{fig1} 
The correlation curve for (Au+Em)-interactions at 10.7 A GeV. 
Points are calculated values, 
lines are fits by (\ref{6})
}
\end{center}
\end{figure}

For our correlation analysis it is possible to use this dependence
visually and
determine the points belonging to the same line and their fit by
(\ref{6}). But
more precisely and more clearly one can investigate the behaviour of the
correlation index $h$ versus the width of pseudorapidity bin $d\eta$
with the help of a 3-point fitting procedure.
Those three points were: the first has been subsequently conferred by
obtained
Hurst index and other two ones are neighboring points.
For example, $h$ corresponding to $k'= 1024$ $(ln(k')= 6.931, d\eta = 0.5)$ in the
Fig.1 has been found, using (\ref{6}) and with following  a fitting
procedure for
points $k'= 512$ $(ln(k')= 6.238)$, $k'= 1024$ $(ln(k')= 6.931)$,
$k= 2048$
$(ln(k')= 7.625)$ from the Fig.1.

\section {Data}

The above analysis procedure has been applied to three samples of events

1. Experimental data of the $ ^{32}S$ 200 A GeV (264 events) and the
   $ ^{197}Au$ 10.7 A GeV (315 events) interactions with emulsion nuclei.
NIKFI BR-2 stacks of nuclear emulsions, sizes of $(10 \times 20 \times 0.06)$
$cm^3$ have been exposed horizontally
to the 10.7 A GeV  $ ^{197} Au$  beam at the BNL/AGS and
the 200 A GeV $ ^{32}S$ at the CERN/SPS. Details about scanning and
measurements are given elsewhere \cite{au},\cite{s}.

2. "Data" obtained by FRITIOF 7.02 model \cite{fr}, of the
$ ^{197}Au$ (10.7 A GeV, 315 "events") and $ ^{32} S$ (200 A GeV, 264 "events")
 interactions with emulsion nuclei.

3. A random process simulated by the HRNDM program, included in the
programming package HBOOK \cite{hb}. Multiplicities and pseudorapidity
distributions for this process are similar to the experimental data
($Au+Em$ at 10.7 A GeV, 315 "events" and $S+Em$ at 200 A GeV, 264 "events").

\section {Results and discussion}

It should be noted that the $\xi_i$ distributions for all data samples are similar (at least
formally) to themselves.
At the same time, the experimentally observed dependence of the
correlation
index on the size of the interval $d\eta$ is different from both HRNDM
and
FRITIOF events (see Fig. 2).

\begin{figure}[tbh]
\begin{center}
\includegraphics*[width=0.48\textwidth,angle=0,clip]{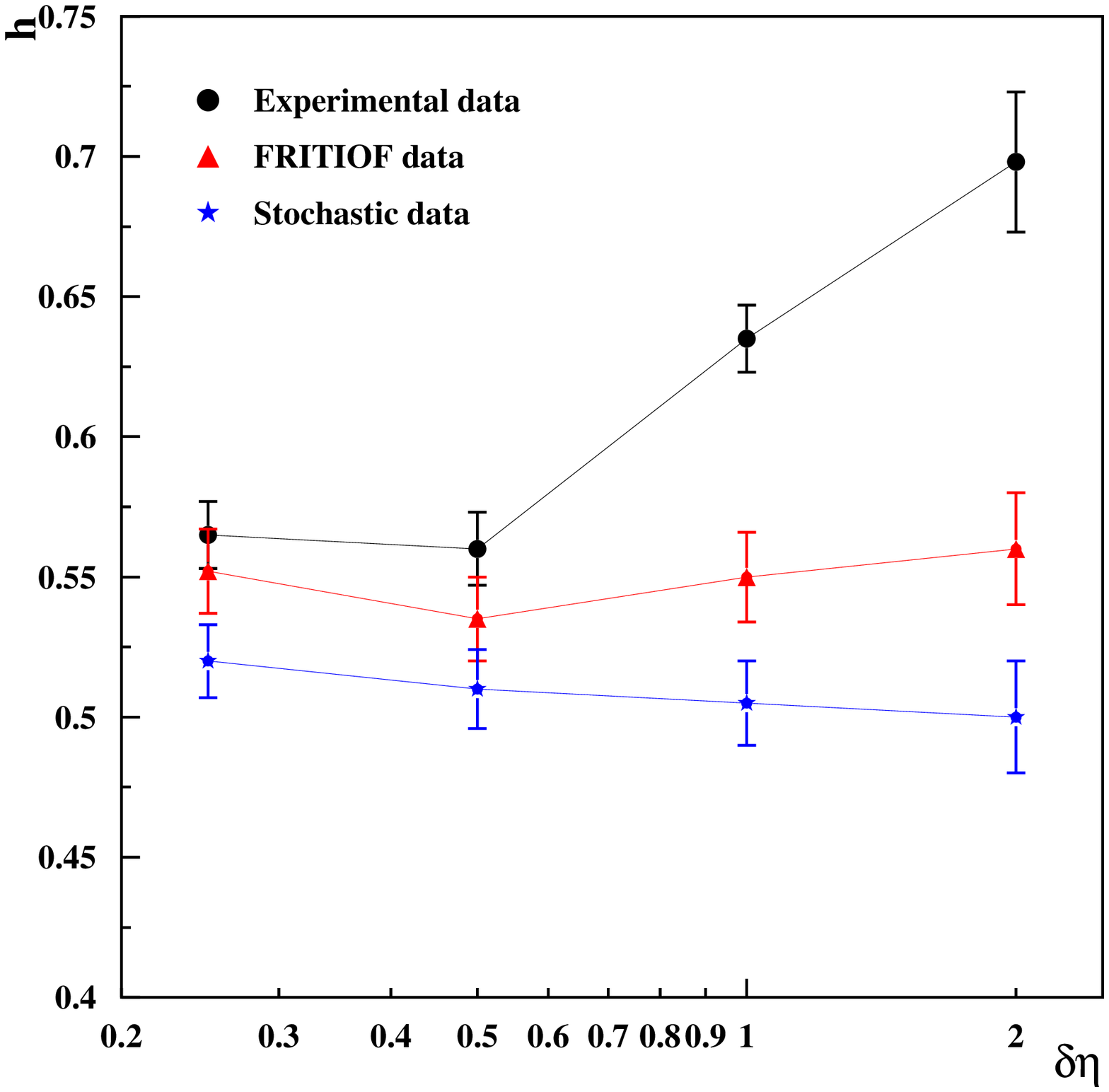}
\includegraphics*[width=0.48\textwidth,angle=0,clip]{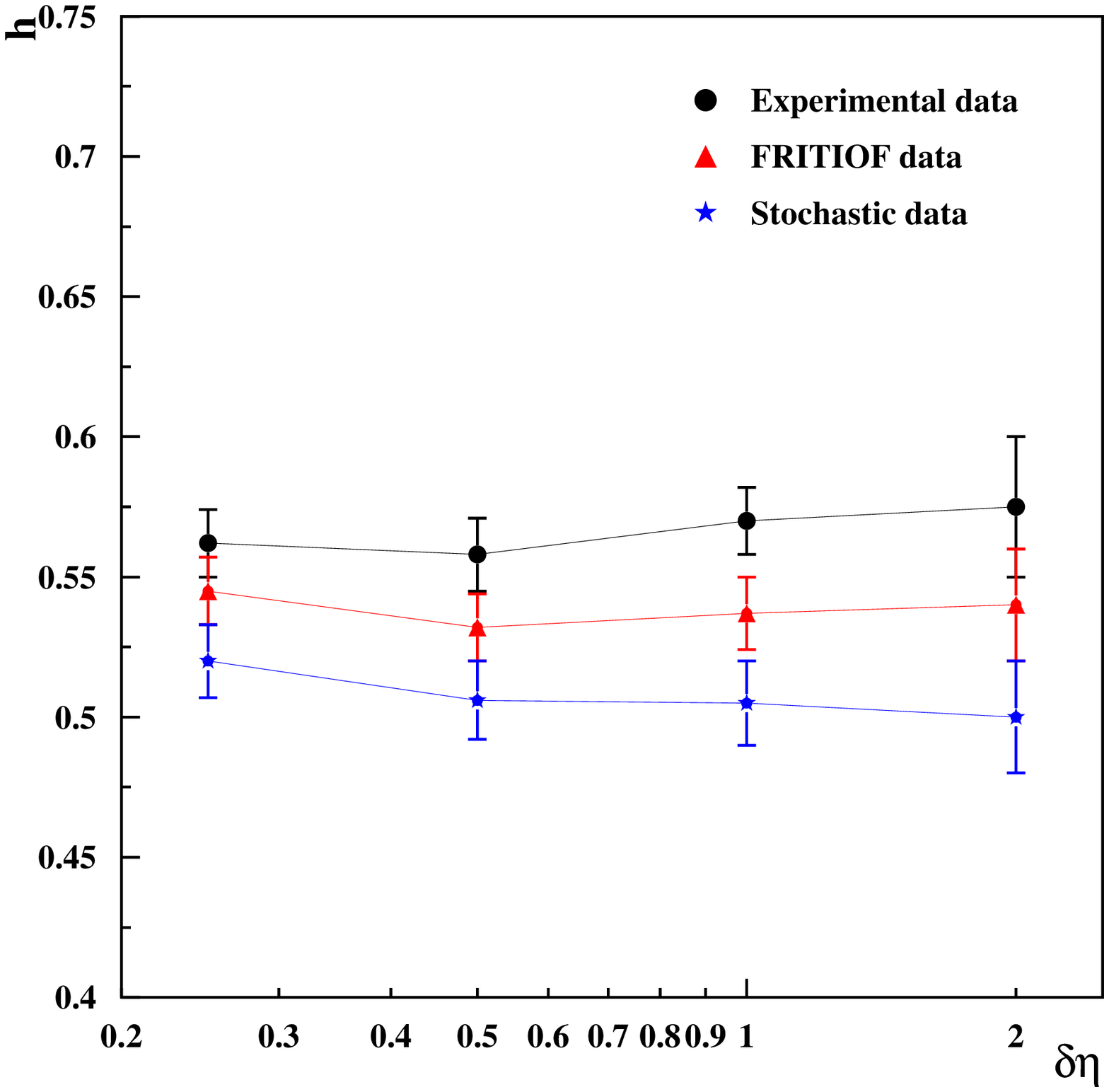}
\caption{\label{fig3} 
The value of Hurst index $h$ versus a width of pseudorapidity bin
$d\eta$ for data of Au+Em 10.7 A GeV (left) and S+Em 200 A GeV (right)
}
\end{center}
\end{figure}

As it is seems from Fig.2 the most obvious difference is evident at large $d\eta$ for (Au+Em)
interactions at 10.7 A GeV, where a significant increase of the
correlation
index is observed, thus indicating stronger correlation "force" \cite{leb}.

A detailed analysis of individual events has shown that all experimental
events
can be conditionally divided into two classes. In Fig. 3 correlation curves for two events from 
two different classes, is presented.   

\begin{figure}[tbh]
\begin{center}
\includegraphics*[width=0.48\textwidth,angle=0,clip]{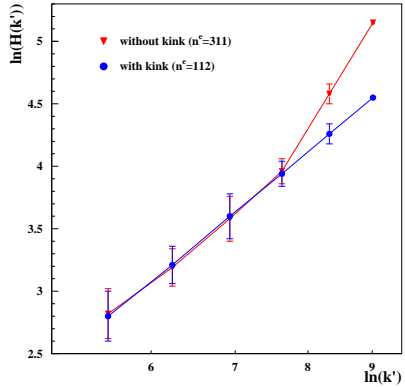}
\caption{\label{fig4} 
The correlation curve for two of events of (Au+Em)-interacions at 10.7 A
GeV
}
\end{center}
\end{figure}

Firstly, the
events
without a "kink" of the correlation curve correspond to random
pseudorapidity
distribution of secondary particles. In these events fluctuations are
essentially statistical, connected with the finite number of particles
in the
event. Secondly, the events "with kink" of the $ln H(k')$ versus $ln k'$
correlation curve, in which we do observe significant multiparticle
correlations at $\delta\eta\ge 1$.

The event selection was made by the average correlation index
\begin{equation}
h_{ev}= \frac{1}{i_{max}-1}\sum_{i= 1}^{i_{max}}
\frac{ln(H(k_i))-ln(H(k_{i-1}))}{ln(k_i)-ln(k_{i-1})}, \; \; (k_i= k/2^i
)
\end{equation}
which was determined by a power-law fit (\ref{6}) to the correlation
curve
curve $ln H (k') $ versus $ln k'$.
If $h_{ev}$ was greater than 0.62, then the event was related to the
group of
events with kink. We can analyze the more correlated events by
increasing
the criterion of selection up to $ h_{ev} \ge 0.7$ or $ h_{ev} \ge
0.8$. In Fig.4 the curves of $ln H(k')$ versus $ln k'$ for 0.62
criterion are shown.

\begin{figure}[tbh]
\begin{center}
\includegraphics*[width=0.48\textwidth,angle=0,clip]{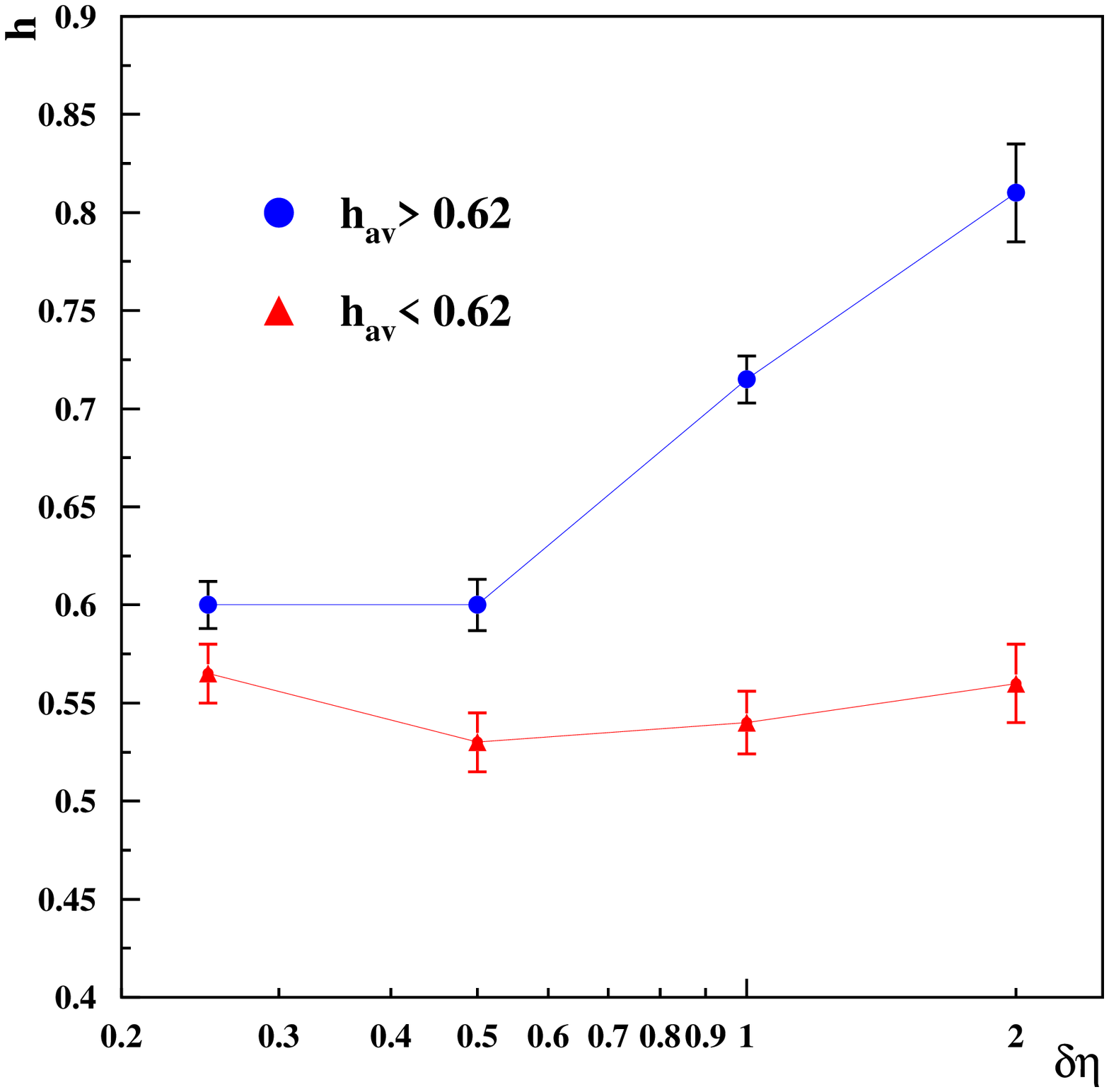}
\includegraphics*[width=0.48\textwidth,angle=0,clip]{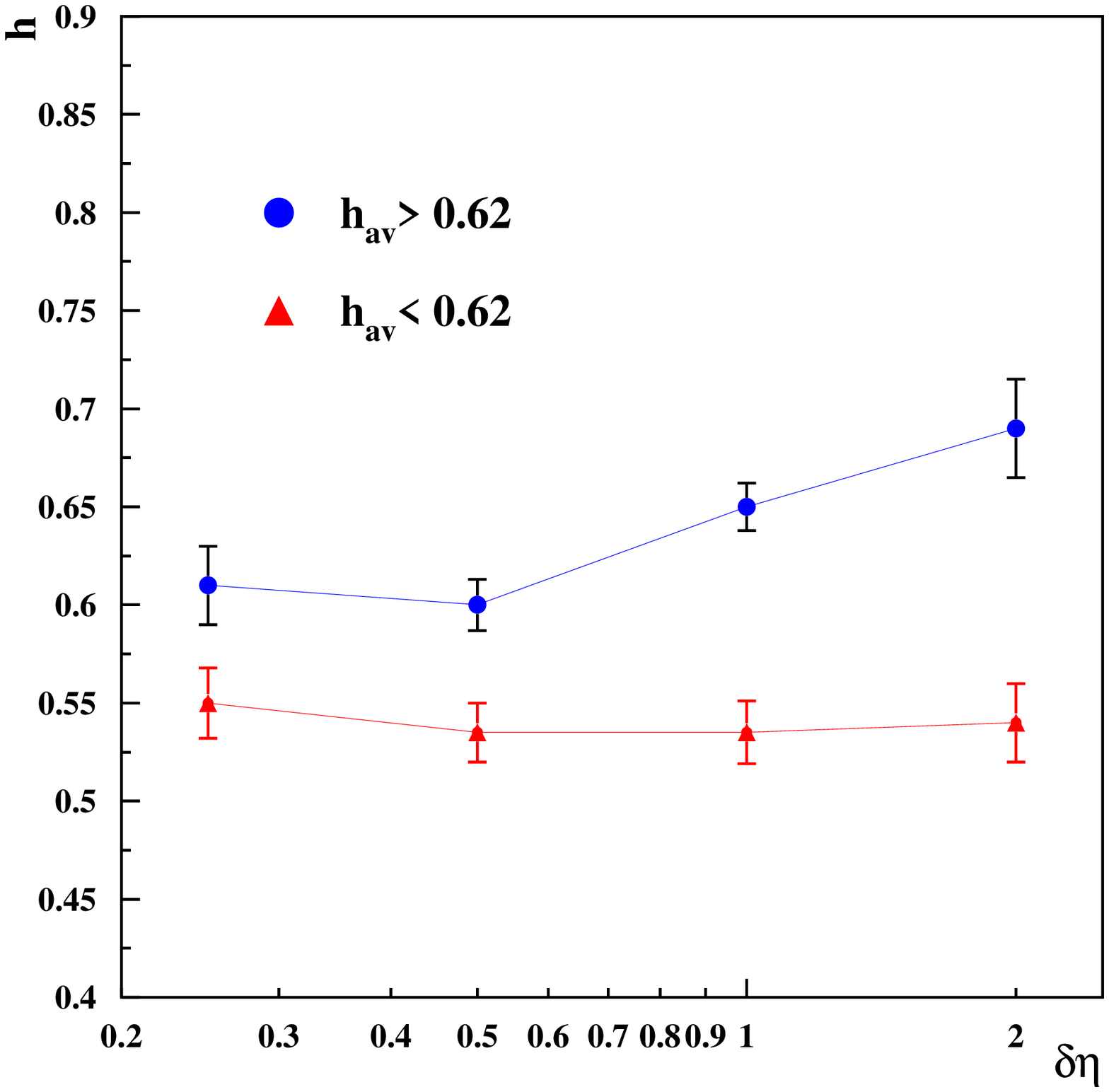}
\caption{\label{fig5} 
The correlation curve for two of events of (Au+Em)-interacions at 10.7 A
GeV
The value of Hurst index $h$ versus a width of pseudorapidity bin
$d\eta$
for the events with $h_{ev}\ge 0.62$ and the events with $h_{ev} < 0.62$ 
from experimental data of Au+Em 10.7 A GeV  (left) and S+Em 200 A GeV (right).
}
\end{center}
\end{figure}

The behaviour of these without kink curves are similar to the
FRITIOF events.

It is interesting to note that the relative number of correlated (S+Em)-events
was considerably lower than for (Au+Em) interactions. Namely, we
have 187 events with $h_{ev}>0.62$ for gold and only 97 ones for sulfur.
The (S+Em)-interactions with very strong correlations (with $h_{ev}>0.8$) were
absent at all. In the case of (Au+Em)-collisions the number of such events
was 16.

To study possible distinctions in the mechanism of formation of a final
condition for the two types of interactions (with kink and without kink
of the
correlation curve), we have analysed the behaviour of fragments of the
projectile and the target nuclei.
Thus, most difference is revealed for the $N_h$-distributions.
$h$-particles are called a sum of $b$- and $g$-particles
($N_h= n_b +
n_g$). According to the existing photoemulsion experiments criteria the
$b$-particles are fragments of target nucleus with kinetic energy
$E_{kin}<26 A
MeV$, $g$-particles are recoil protons with $26<E_{kin}<400 MeV$.
\footnote{It should be noted that h-particles and Hurst index (h)
have not any connection between themselves. The signs have coincided
by accident.}
As it is seen from Fig. 5 a large part of events with kink, as opposed to
the interactions without kink,
proceed with a complete (or almost complete) disintegration of the target
nucleus (peak in area $N_h= 0$). 

\begin{figure}[tbh]
\begin{center}
\includegraphics*[width=0.48\textwidth,angle=0,clip]{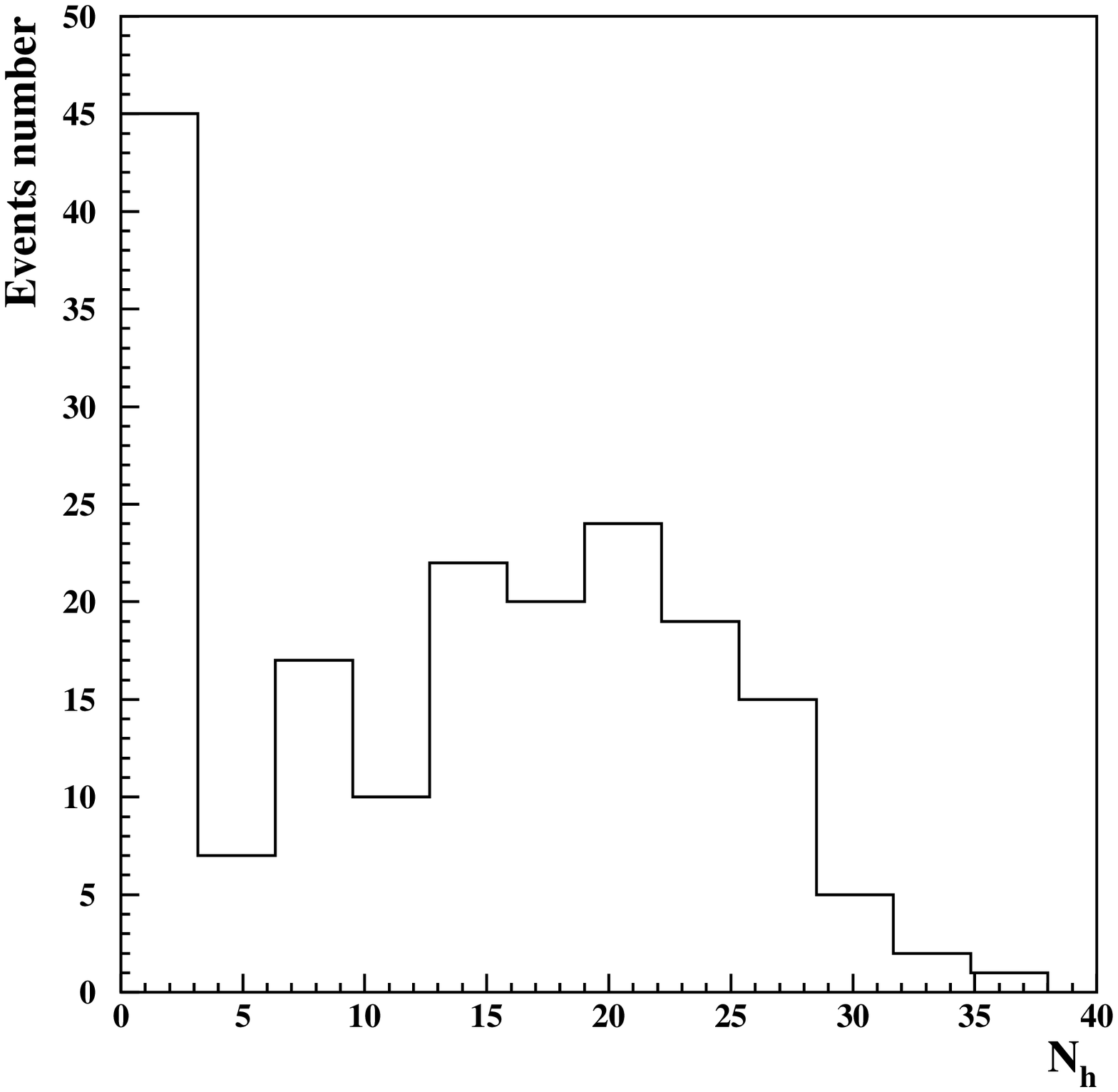}
\includegraphics*[width=0.48\textwidth,angle=0,clip]{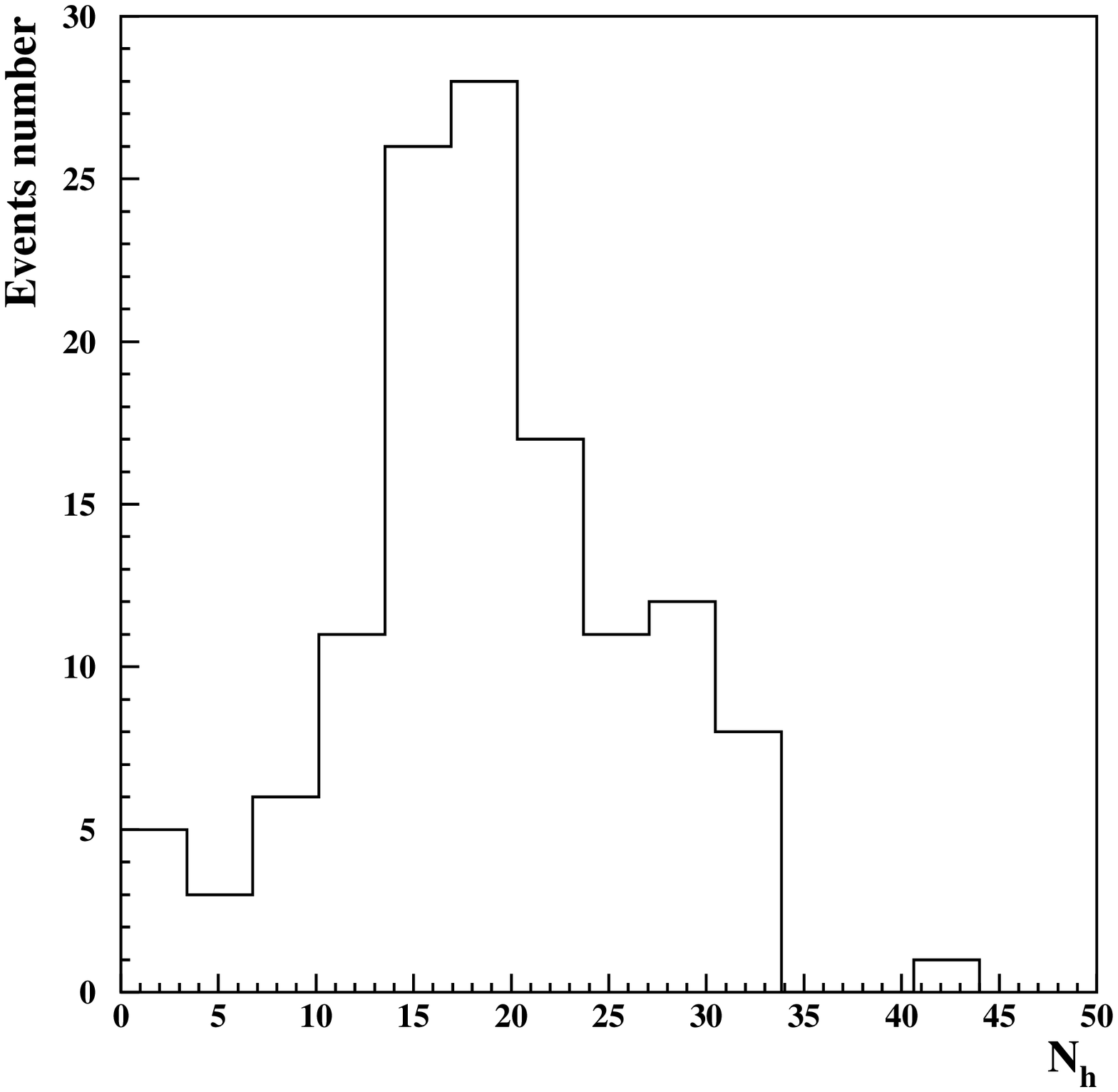}
\caption{\label{fig6} 
The $N_h$-distribution for data of Au+Em (10.7 A GeV) for the events with $h_{ev}\ge 0.62$ (left) and 
for the events with $h_{ev} < 0.62$  (right)
}
\end{center}
\end{figure}

Most of these kink events (events with very
strong correlations) corresponds to such events (Fig. 6left).

\begin{figure}[tbh]
\begin{center}
\includegraphics*[width=0.48\textwidth,angle=0,clip]{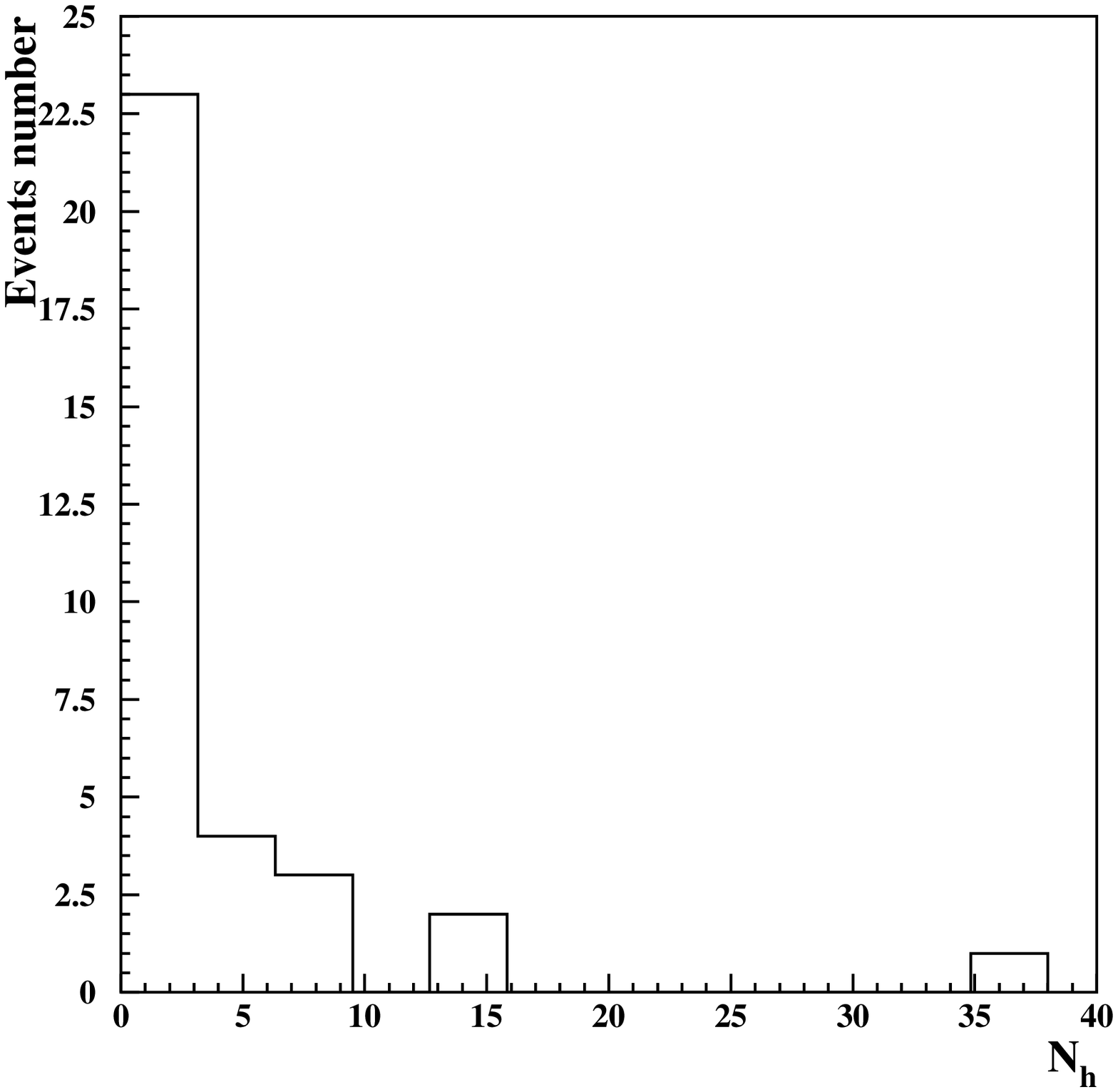}
\includegraphics*[width=0.48\textwidth,angle=0,clip]{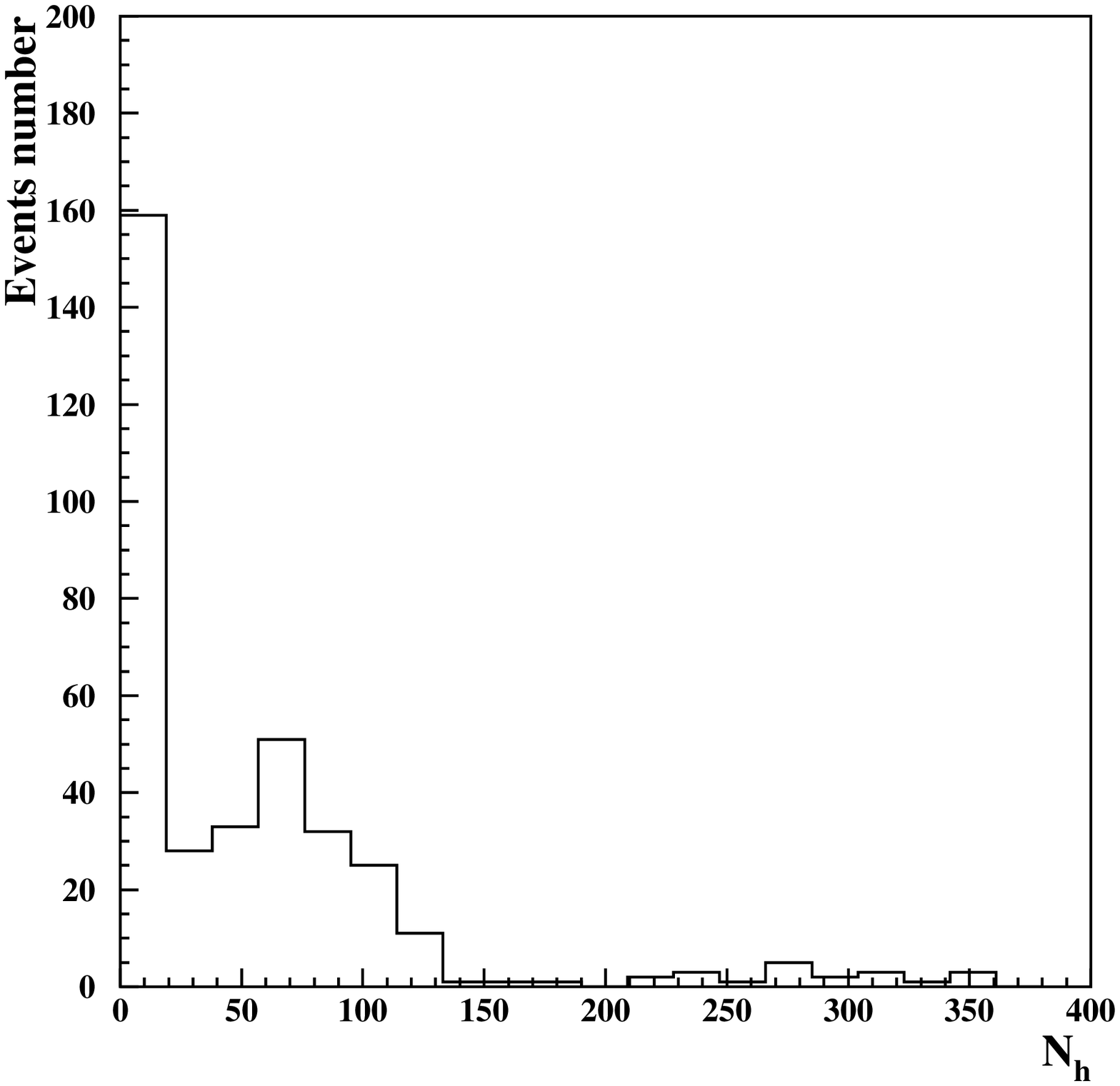}
\caption{\label{fig7} 
The $N_h$-distribution for data of Au+Em (10.7 A GeV) for the events with $h_{ev}\ge 0.8$ (left) and 
the $N_s$-distribution for data of Au+Em (10.7 A GeV) for the events with $n_{b}= 0$ (right)
}
\end{center}
\end{figure}

The photoemulsion consist of Hydrogen(39.2\%), nuclei of $CNO$-group(35.3\%)
and AgBr nuclei (25.5\%). In order to know what interactions give main
contribution in peak of $N_h=0$ we have analysed
$n_s$ distribution
\footnote{$s$-particles are secondary ones with $E_{kin}>400 A MeV$.}
for events with $n_b=0$ (Fig. 6right).

The first peak in this
distribution at $n_s < 30$ corresponds to peripheral interactions of
gold with
photoemulsion nuclei and central interactions, $(p+Au)$. The second peak
at $30
< n_s < 200$, is a result of central (since $n_s$ are large)
interactions of
light nuclei (CNO group) and nuclei of gold. The peak in the area of
$n_s>200$
corresponds to central collisions of $(Ag + Au)$ and $(Br + Au)$.

Hence, by analysing events with $n_s > 200$, we study interactions of
heavy
nuclei of $Ag$ and $Br$ with nuclei of gold. We expect that greatest
correlations in such interactions. However, as it can be seen from Fig.
7left, the
peak in area of $N_h= 0$, corresponding to the greatest values of a
correlation
index, has disappeared!

\begin{figure}[htb]
\begin{center}
\includegraphics*[width=0.48\textwidth,angle=0,clip]{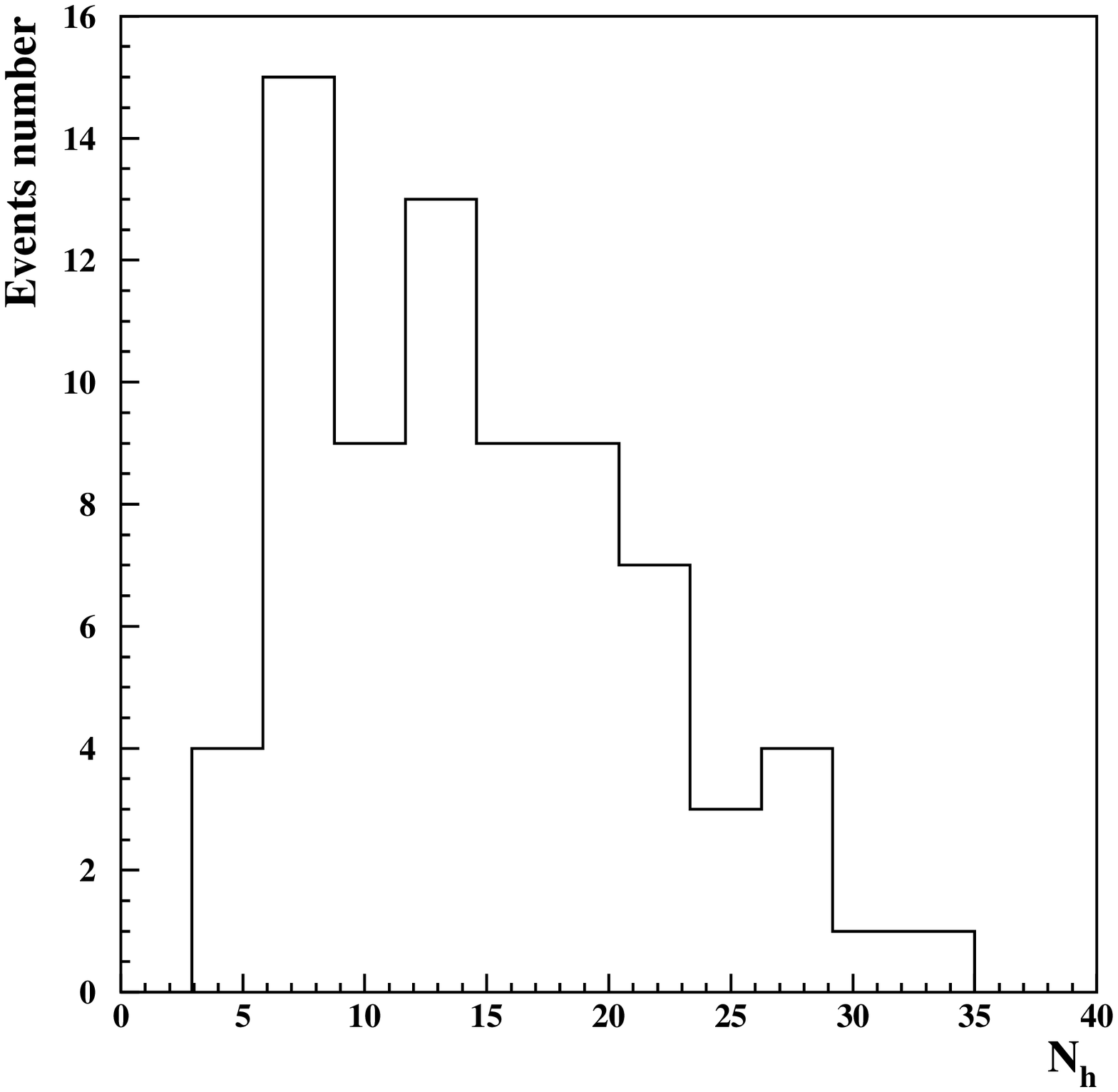}
\includegraphics*[width=0.48\textwidth,angle=0,clip]{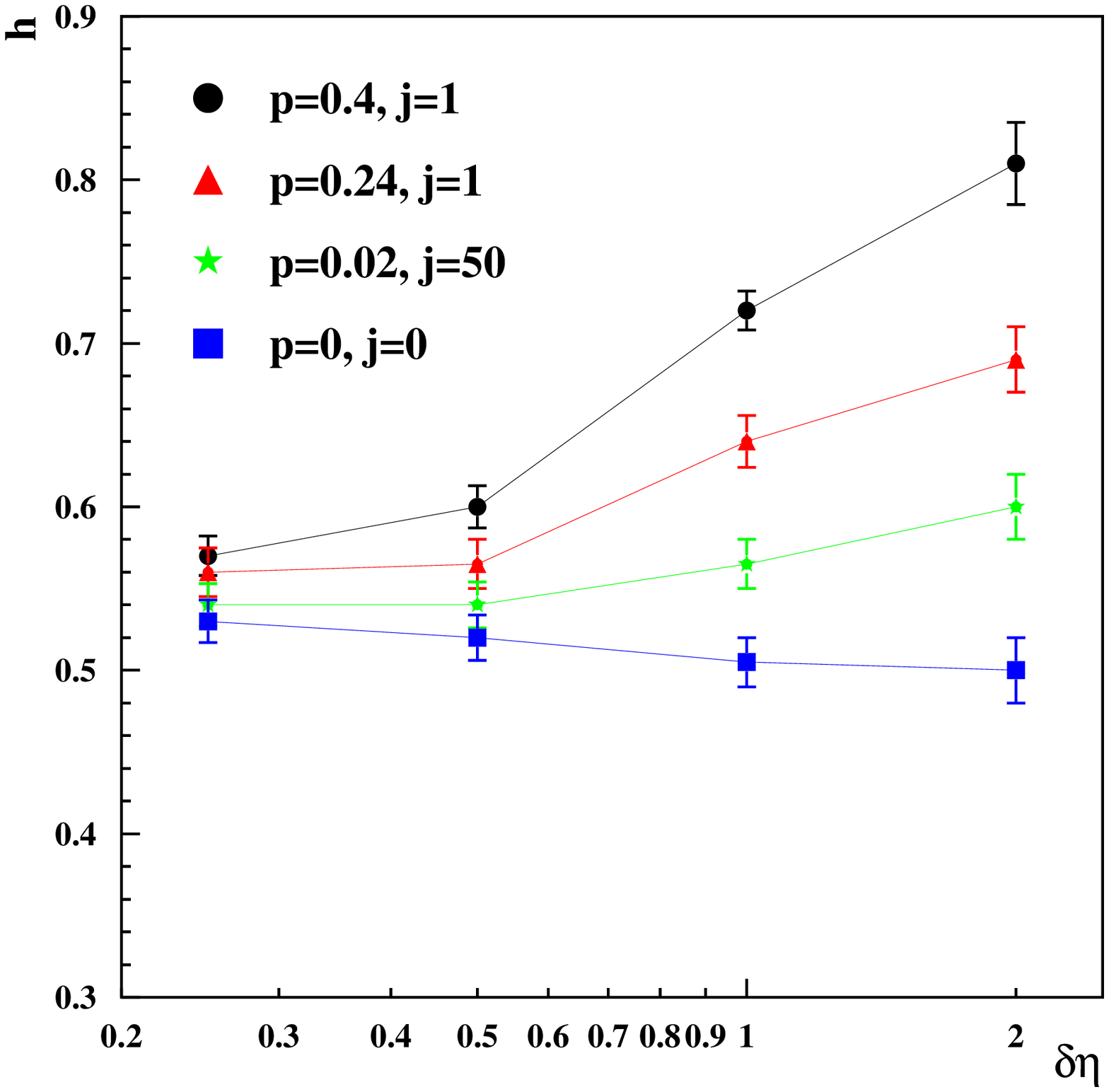}
\caption{\label{fig8} 
The $N_h$-distribution for data of Au+Em (10.7 A GeV) for the events with $n_{s}\ge 200$
 (left) and the value of Hurst index $h$ versus a width of pseudorapidity bin
$d\eta$ for modelled data with $RN=1$ and different p and j.
}
\end{center}
\end{figure}

Therefore, the events which give the main contribution to the peak in
the area
of $N_h \sim 0$ and which have great kink of correlation curve, correspond
to
central interactions of $Au$ and light nuclei of $CNO$ group.

It is possible that there are several probable reasons for the existence of
such large correlations in multiparticle production processes.
But, nevertheless, at present we know only one way when
similar behaviour (to experimental one) of the correlation curve can be
obtained.
As result of interactions of two particles at high energy
there is formed some large claster (or several clasters), which then
disintegrates into a plenty of secondary particles
in the interval of $\delta \eta < 1$.

This process was simulated with the help of the programme "HRNDM",
included into the software package of "HBOOK" \cite {hb}.
We simulated stochastic events with pseudorapidity distributions similar
to experimental data.
We have designate: a total number of generated "particles" in an event as $n^e$,
probability of production of claster as $p$ and number of the
clasters as $j$ ($0\le pj\le 1$).
Each simulated "particle" can be disintegrated (with probability $p j$)
into $p n^e$-particles with pseudorapidity $\eta_l=\eta_m + f_l$,
$1\le l\le p n^e$,
where $f_l$ is random function, which changes with uniform probability in
region [-$RN$,+$RN$],
$\eta_m$ is pseudorapidity of the "parent particle".

By varying $p$, $j$ and $RN$, we can define curves most precisely
reproducing the behaviour of our experimental data.

From Fig.7right it is seen that
the more dimension of the compound system, the more "kink" of the correlation
curve will be displayed.

Moreover, such the large kink of correlation curve, which observed for
experimental events, cannot be explaned by two-particle
disintegrations. So, the curve with j=50 and p=0.02 in Fig.7right corresponds to process with
maximal two-particle correlations (all secondary particles are produced
as result of two-particles disintegrations). In this case
the correlation index reached only 0.62. In order to obtain the values $h$
which observed for experimental data, it is necessary to assume that the
compaund system, which produces $\sim 40$ \% (for Au+Em) and $\sim 24$\%
(for S+Em) of secondary particles, is formed.

It should be noted that for search of the most interesting effects
as quark-gluon plasma, for example, in CERN experiments as NA44, NA49,
WA98 etc. symmetric collisions of heavy ion as Pb+Pb (or Au+Au) are used.
If our results will be confirmed by other experimental data (for instance,
in interactions of $Au$ (or $Pb$) and nuclei of $CNO$ group at other (more or
less than 10.7 a GeV) energies) then it will be possible
declare that probability of exhibition of QGP in asymmetric interaction is
more than that in symmetric ones.

To clear this point it is possible by primitive model \footnote{It should
be noted that the process of nuclei interaction at high energies is more
complicated one. Nevertheless, the model allow to realize why largest
correlations should be exhibited in asymmetric collisions instead of
symmetric ones".}. Supposing that a
nucleon is an elastic ball, inside which quarks are disposed, and nucleus
is some system consisting the balls, which are glued together. At the
moment of collision first balls, coming into contact, are squeezed by other
balls, which push them from behind and in front. And so, a real
energy of interaction (at this "region") is significantly more than an energy
per nucleon (some cumulative effect). A pressure (at this "region") is
enormous. Volumes of the "interacting balls" is getting significantly
diminished (is squeezed). As result a nucleon "envelope" can burst
and quarks (from the nucleon) will be in quasifree state inside dense
"encirclement" of  squeezed (but not bursting) neighbouring nucleons.
This region can exist and be expanded till large nucleus are not split
(when "force of glue" will be less than "breaking force of the wedged
(smaller)
nucleus"). If an interaction is less central (almost peripheral) one,
then quarks region and time of its existence will be less, because of that
the probability to chop off smaller "scrap" is more than that to split
nucleus to "fifty-fifty".
In case of collision of symmetric nuclei the quarks region
can appear also, but its dimensions will be still less, because
dense "encirclement", which is capable to retain (inside itself) quarks, is
absent (it is similar
to peripheral interaction of asymmetric nuclei).

\section{Conclusion}

The analysis of pseudorapidity correlations in 10.7 A Gev $^{197}Au$ and
200 A GeV $^{32}S$ interactions with emulsion nuclei has been carried
out
by the normalized range method.
This analysis has revealed events with large multiparticle correlations
(with large kink of correlation curve).
We assume that the main cause of this behaviour of the correlation curve
is
a formation of large compound system, which then disintegrates into a
large
number of secondary particles in the interval of $\eta \pm \delta \eta$
($\delta \eta < 1$).
The most significant pseudorapidity correlations are observed in the
central interactions of nuclei, which have considerable difference in
volume
(in nuclear weight, charge, etc.) at rather "low" energies
(the nuclei of Au and CNO group at 10.7 A GeV).

It would be interesting to see whether a similar effect is observed in
interactions of other nuclei (for example, lead and nuclei of CNO group)
and whether it depends on the interaction energy (if the effect will be
increase at energy of collision more than 10.7 A GeV).


\begin{thebibliography}{99}
\bibitem{drem}
I.M. Dremin, A.V. Leonidov
Phys.Usp.53 1123-1149, 2011; arXiv:1006.4603v2 [nucl-th] 2010 
\bibitem{leb}
I.A.Lebedev, B.G.Shaikhatdenov
J.Phys.G:Nucl.Part.Phys. 23 (1997) 637
\bibitem{h}
H.E. Hurst, R.P. Black, Y.M. Simaika (1965), Long-Term Storage: An
Experemental Study (Constable, London)
\bibitem{f}
J. Feder "Fractals", Plenum Press, New York, 1988,
\bibitem{au}
M.I.Adamovich et al. Phys.Lett. B352 (1995) 1472
\bibitem{s}
M.I.Adamovich et al. Phys.Lett. B227 (1989) 285
\bibitem{fr}
B.Nilsson-Almqvist, E.Stenlund, Comut.Phys.Commun. 43 (1987) 387,
\bibitem{hb}
R.Brun, D.Lienart,  CERN computer centre program library long write-up.
\end{thebibliography}
\end{document}